\documentclass[]{aa}
\usepackage{graphicx}
\begin{document}
\title{Intergalactic HeII absorption towards QSO~1157+3143 \thanks{Based on observations with the NASA/ESO Hubble Space Telescope, obtained at the Space Telescope Science Institute, which is operated by Aura, Inc. under NASA contract NAS 5 - 26555. Based on observations from the German-Spanish Astronomical Center, Calar Alto, and based on observations at the Keck Observatory.}
}
\subtitle{}

\author{D. Reimers 
        \inst{1}
        \and 
	C. Fechner\inst{1}
	\and 
	H.-J. Hagen\inst{1}
	\and 
	P. Jakobsen\inst{2} 
	\and 
	D. Tytler\inst{3}
	\and 
	D. Kirkman\inst{3}
       }

\offprints{}

\institute{Hamburger Sternwarte, Universit\"at Hamburg,
   Gojenbergsweg 112, D-21029 Hamburg, Germany
\and 
  Astrophysics Missions Division, RSSD, European Space
  Agency, ESTEC, 2200 AG Noordwijk, The Netherlands
\and 
  Center for Astrophysics and Space Sciences,
  University of California, San Diego, MS 0424, La Jolla,
  CA 92093-0424, USA
}

\date{received 4 May 2005/ accepted 30 June 2005}

\abstract{
We report the discovery of a further line of sight allowing detection
of \ion{He}{ii} Ly$\alpha$ absorption by the intergalactic medium. 
A HST/STIS survey of 32 bright $z \sim 3$ quasars selected from the Hamburg Quasar Surveys yielded one detection toward QSO~1157+3143 ($z \sim 3, B \simeq 17$). 
A 10 orbit follow-up spectrum reveals a UV spectrum significantly supressed by two intervening Lyman limit systems at $z=2.77$ and $2.94$, but with the
continuum flux recovering sufficiently shortward of $\sim 1700\,\mathrm{\AA}$ to allow study of the \ion{He}{ii} absorption spectrum in the redshift range $2.75 \leq z \leq 2.97$. 
The absorption is characterized by alternating voids and
dense filament structures seen in both \ion{He}{ii} and \ion{H}{i}. 
Attempts to model the \ion{He}{ii} opacity in terms of \ion{H}{i} Ly$\alpha$ forest absorption are not successful in the voids, suggesting that \ion{He}{ii} reionization is not complete between $z=2.77$ and $2.97$ or that an optically thin Lyman limit system with $z \approx 0.3$ is responsible for the additional opacity.

}   

\maketitle
%

\section{Introduction}
Among the important results in observational cosmology in the past decade is the detection of  intergalactic \ion{He}{ii} 304\,\AA\, absorption in a few lines of sight, e.g. Q~0302-003 (Jakobsen et al. \cite{jak}), PKS~1935-6914 (Tytler et al. \cite{tyt}), HS~1700+6416 (Davidsen et al. \cite{dav}), and HE~2347-4342 (Reimers et al. \cite{rei1997}). In particular, it has been shown that for $z \geq 3$ the \ion{He}{ii} absorption is optically thick, i.e. no remnant flux has been observed, while for $z < 2.9$
the \ion{He}{ii} opacity becomes patchy (first seen in HE~2347-4342) which provided
strong evidence for a delayed cosmic \ion{He}{ii} reionisation phase around
redshift 3 (Reimers et al. \cite{rei1997}). Tentative independent observational evidence for the
reality of the delayed \ion{He}{ii} reionisation phase comes from the observation of a sudden increase in the line widths of Ly$\alpha$ forest lines in the
redshift range between 3.5 and 3.0 interpreted as reheating of the IGM
due to \ion{He}{ii} reionisation (Ricotti et al. \cite{ric}, Schaye et al. \cite{sch}, Theuns et al. \cite{the}). 
Thus observations indicate that \ion{He}{ii} reionisation started around redshift 3.5 and was complete around $z = 2.8$ where the last surviving patches of \ion{He}{ii} 
between the \ion{He}{iii} Str\"omgren sphere finally become optically thin.
This final ``patchy'' phase, the transition between continuous, ``black''
absorption ($z > 3$) and the reionized phase, the \ion{He}{ii} Ly$\alpha$ forest,
which has been resolved for the first time by FUSE observations of
HE~2347-4342 for $z < 2.8$ (Kriss et al. \cite{kri}), is actually not well
covered by existing observations. It is seen only in Q~0302-003
($2.8 < z < 3.28$) and in HE~2347-4342 ($z < 2.89$).
In this paper we report on \ion{He}{ii} observations of a third case in the line of sight of \object{HS~1157+3143} = \object{CSO118} ($z \simeq 3$, $\alpha(1950.0) = 11^{\mathrm{h}} 57^{\mathrm{m}} 32\fs3$, $\delta(1950.0) = +31\degr 43\arcmin 12\arcsec$), hereinafter called \object{QSO 1157+3143}, which covers the redshift range $z = 3.0$ to $2.77$. The clear line of sight to this quasar has been discovered in HST/STIS survey programs on altogether 32 bright
QSOs from the Hamburg objetive prism QSO surveys. We will show that the
QSO 1157+3143 observations confirm the above outlined picture of \ion{He}{ii}
reionisation and show in addition for the first time large scale structure
in the \ion{He}{ii} opacity.

\section{Observations}
\begin{figure}
  \centering
  \resizebox{\hsize}{!}{\includegraphics[bb=35 505 530 770,clip=]{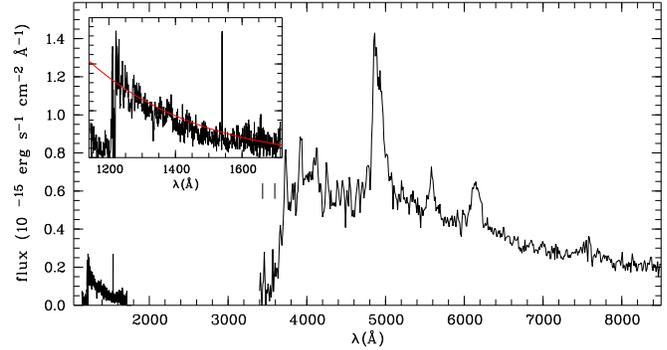}}
  \caption{
    Observed energy distribution of QSO 1157+3143 in the optical (from CAFOS) and in the UV (STIS/HST). The inset shows the model continuum predicted from a QSO power law spectrum and a Lyman limit system at $z = 2.77$ with $\log N_{\ion{H}{i}} = 18.3$. Vertical bars show the locations of the Lyman limit systems at $z = 2.77$ and $z = 2.94$.
  }
  \label{fig1}
\end{figure}

\begin{figure}
  \centering
  \resizebox{\hsize}{!}{\includegraphics[bb=45 30 510 260,clip=]{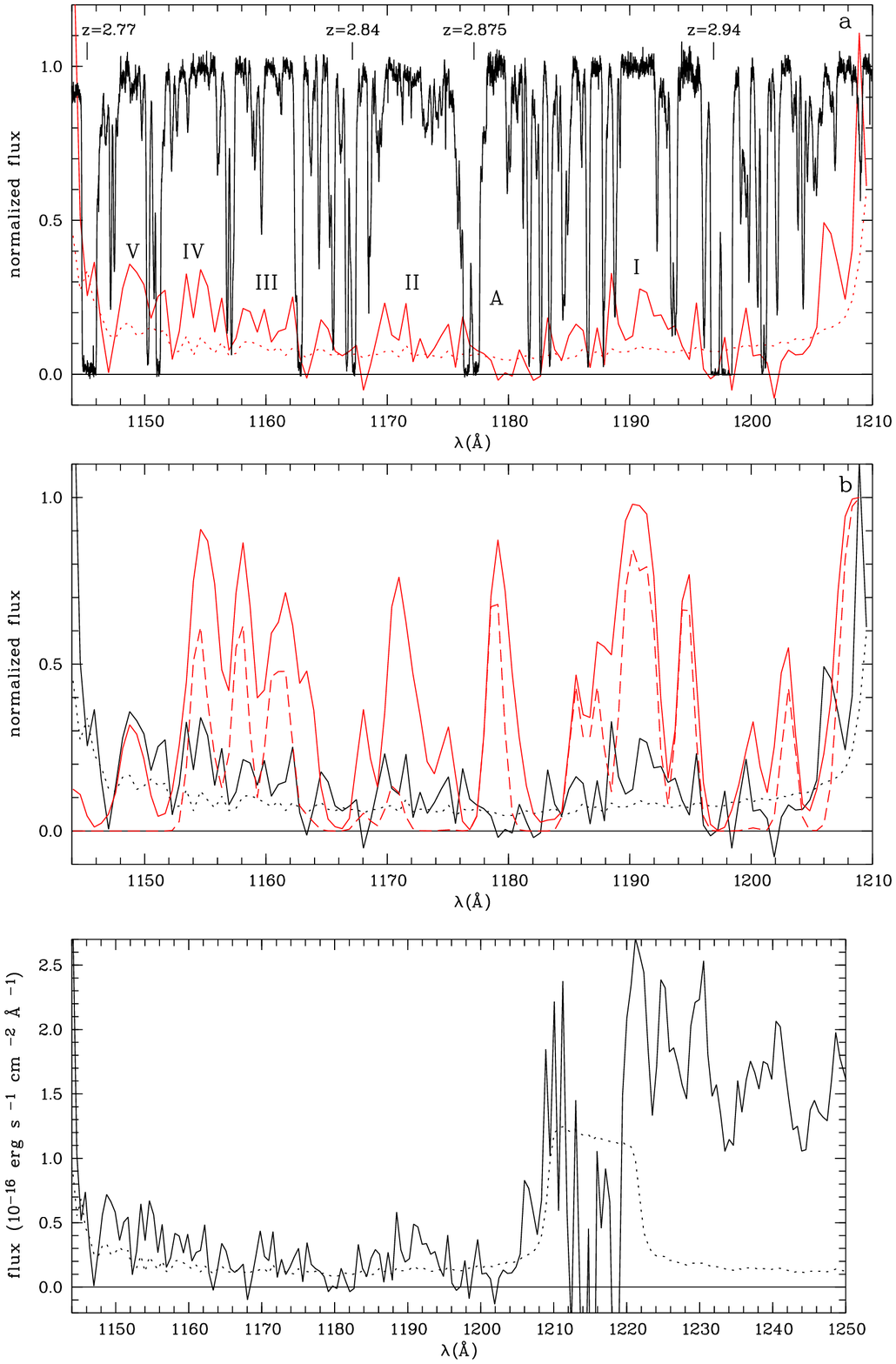}}
  \caption{\ion{He}{ii} 304\,\AA\, absorption in QSO 1157+3143.
  }
  \label{fig3}
\end{figure}

The small number of \ion{He}{ii} detections (altogether 6 including the present
one) is due to the fact that the \ion{He}{ii} 304\,\AA\, line is accessible only in
high-redshift QSOs ($z \geq 2.8$ with HST) and that the chances to find a
clear line of sight is only of the order of a few percent at redshift 3.
The \ion{He}{ii} 304\,\AA\, line lies deep in the Lyman continuum where the
continuum fluxes of the vast majority of high redshift quasars are severely
absorbed by intervening neutral hydrogen contained in the Ly$\alpha$ forest
and especially the denser Lyman limit systems encountered at all redshifts
along the line of sight (M{\o}ller \& Jakobsen \cite{mol}). This opacity --
when combined with typical brightnesses of quasars at redshifts of the
order of 3 needed to reach the \ion{He}{ii} 304\,\AA\, transition -- effectively
confines \ion{He}{ii} absorption measurements to continuum sources having fluxes
of order $f_{\lambda} = 10^{-15}\,\mathrm{erg\,s}^{-1}\,\mathrm{cm}^{-2}\,\mathrm{\AA}^{-1}$
(Picard \& Jakobsen \cite{pic}).
We have therefore in the late 1980s embarqued on a basically all --
extragalactic -- sky survey for bright quasars based on objective prism
Schmidt plates taken with the Calar Alto and ESO Schmidt-telescopes
(Hagen et al. \cite{hag}, Wisotzki et al. \cite{wis}). UV follow-up spectroscopy on the brightest
targets ($B < 16.5$) which appeared to be free from Lyman limit and strong
metal line absorption has been conducted subsequently with IUE, HST and
FUSE in several survey programs (for a more detailed description cf.
Reimers \& K\"ohler \cite{rei1997}). The two brightest objects
in which the \ion{He}{ii} Ly$\alpha$ forest has been resolved with FUSE,
HE~2347-4342 (Kriss et al. \cite{kri}) and HS~1700+6416 (Reimers et al. \cite{rei2004}) have
been discovered already with IUE to have 'clear' lines of sight
(Reimers et al. \cite{rei},\cite{rei1997}). 
QSO 1157+3143 was discovered to have a transparent line of sight in
the course of 2 HST survey programs (GO 7471 and GO 8287) on altogether
32 bright targets from the Hamburg surveys. Only one successful detection,
namely QSO 1157+3143, resulted from this survey. A success rate of $\sim$3\,\%
is roughly consistent with the prediction of Picard \& Jakobsen (\cite{pic})
and with the result of a similar program on fainter QSOs from the SDSS
(Zheng et al. \cite{zhe2004a}). 
QSO 1157+3143 was first detected in the course of the CASE Survey
as CSO118 (Everett \& Wagner \cite{eve}). With $f_{5000\,\mathrm{\AA}} \simeq 0.5\cdot 10^{-15}\,\mathrm{erg\,cm}^{-2}\,\mathrm{s}^{-1}\,\mathrm{\AA}^{-1}$ it is only a factor
of 2 fainter than HE~2347-4342 (Fig. \ref{fig1}).
However, its UV flux
is largely absorbed by two intervening Lyman limit systems at $z = 2.94$
and $z = 2.77$ (Kirkman \& Tytler \cite{kir}).
The rest frame redshift of the QSO is difficult to measure. Everett \& Wagner (\cite{eve}) give $z = 2.97$ from a low resolution spectrum using \ion{C}{iv}
and Ly$\alpha$. The Ly$\alpha$ forest in QSO 1157+3143 begins at
$z = 3.00$ which is probably closer to the truth.
QSO 1157+3143 has been detected to have a non zero flux redward of
the expected position of the onset of \ion{He}{ii} absorption at $\lambda >1220\,\mathrm{\AA}$ in a 1 orbit exposure.
Spectral data of this quasar in the UV range were obtained with the
HST/STIS. The log of observations is given in Table \ref{tab1}.
Since the flux level in the wavelength region with \ion{He}{ii} absorption is
extremely low ($\simeq 0.5\cdot 10^{-16}\,\mathrm{erg\,cm}^{-2}\,\mathrm{s}^{-1}\,\mathrm{\AA}^{-1}$), special care has been taken to determine the background. The main difference to the pipeline reduction was that we determined the background on the co-added 2D frame.
The resulting spectrum is shown in Fig. \ref{fig2}a. 
In Fig. \ref{fig1} we display the HST/STIS spectrum in combination with a flux calibrated low resolution spectrum taken with CAFOS at the Calar Alto 2.2m telescope.
Fig. \ref{fig3} shows an enlarged section of the STIS spectrum.
In addition, QSO 1157+3143 has been observed for a total of 7 hrs
in 1996 March and 1997 January with the HIRES spectrograph on the
W. M. Keck telescope. The resulting spectrum has a resolution of
$7.9\,\mathrm{km\,s}^{-1}$ and a $S/N \sim 40$ per $0.03\,\mathrm{\AA}$ pixel (cf. Kirkman \& Tytler \cite{kir}).

\begin{table*}
  \caption[]{The log of observations.
  }
  \label{tab1}
  \begin{center}
  \begin{tabular}{llclrl}
    \hline
    \noalign{\smallskip}
    Telescope & Spectrograph & $\lambda \lambda$ & Resolution & Exp time & Date \\
    \noalign{\smallskip}
    \hline
    \noalign{\smallskip}
    Calar Alto 2.2m & CAFOS      & $3200-8500$ & 2.0\,\AA    &   600 s & March 4 1997 \\
    HST             & STIS G140L & $1150-1750$ & $\sim 1000$ &  2365 s & Oct 30, 2001 \\
                    & STIS G140L & $1150-1750$ & $\sim 1000$ & 26820 s & July 2003  \\
    Keck            & HIRES      & $3800-6200$ & 38000       &   7 hrs & Dec 96/Jan 97\\
    \noalign{\smallskip}
    \hline
  \end{tabular}
  \end{center}
\end{table*}

\section{Observational results}
In this section we analyse the \ion{He}{ii} absorption seen shortward of 
$1205\,\mathrm{\AA}$ in terms of Ly$\alpha$ forest absorption. The \ion{He}{ii} 304\,\AA\,
forest has been resolved by FUSE in HE~2347-4342 and HS~1700+6416.
In particular we wish to find out whether all \ion{He}{ii} absorption seen is due to
the normal Ly$\alpha$ forest or whether additional ``continuous''
\ion{He}{ii} absorption due to patches of not yet reionized matter are present
in the redshift range 2.75 to 2.97.

\subsection{The continuum redward of 1220 \AA\,}

For an analysis of the \ion{He}{ii} opacity, good knowledge of the QSO
continuum is necessary. Due to the high \ion{He}{ii} opacity for $\lambda < 1210\,\mathrm{\AA}$, the continuum is nowhere detectable directly and has to be
determined by an extrapolation from the continuum redward of 1220\,\AA.
As can be seen from Fig. \ref{fig2}, the UV continuum of the QSO is dominated by
strong Lyman limit absorption around 3600\,\AA\, from which the continuum
recovers thanks to the $\nu^{-2.7}$ dependence of the Lyman continuum
opacity on frequency $\nu$ . We model the QSO continuum in two steps.
First we determine the optical depth of the Lyman continuum absorption
by a comparison with the extrapolated QSO continuum (Fig. \ref{fig1}).
As discussed by Kirkman \& Tytler (\cite{kir}), there are probably two LLS,
one at $z = 2.94$ and one at $2.77$. From the available data it is impossible
to decide unambiguously which one is dominant for the optically thick
Lyman continuum absorption in the UV, since the $z=2.94$ LLS blots out the spectrum below 3600\,\AA, and the $z=2.77$ Lyman lines Ly$\alpha$ to Ly$\gamma$ are all saturated. 
Our guess is that the system
responsible for most of the Lyman continuum absorption is the system 
at $z = 2.77$. It is a complex, multiphase absorption system with
at least 5 components which contains sharp lines from ions like \ion{Si}{ii}, \ion{C}{ii}, or \ion{Si}{iii}, broad lines of \ion{C}{iv}, and even broader lines of \ion{O}{vi} (Kirkman \& Tytler \cite{kir}). However, for the total optical depth of the combined
$z = 2.77 / 2.94$ system, which we determine from the height of the
continuum for $\lambda < 1700\,\mathrm{\AA}$, the exact position of the effective
Lyman limit ($3445$ versus $3589\,\mathrm{\AA}$) is less important than the assumed
run of the QSO continuum shortward of 4000\,\AA\, (cf. Fig. \ref{fig1}).
With an assumed power law QSO continuum $f_{\lambda} \sim \lambda^{-2.0}$
between 4000\,\AA\, and 8000\,\AA, a good fit to the UV continuum between
$\sim 1700\,\mathrm{\AA}$ and $\sim 1250\,\mathrm{\AA}$ yields $\tau_{\mathrm{LL}} \simeq  20$ at $z = 2.77$ or $z = 2.94$ which corresponds to $\log N_{\ion{H}{i}} \simeq 18.3$.

\subsection{The continuum in the \ion{He}{ii} forest}

Having modelled the observed continuum between 1220\,\AA\, and 1700\,\AA\, we assume
that this continuum can be extrapolated to $\lambda < 1200\,\mathrm{\AA}$. We
neglect galactic extinction which is low given to the galactic latitude
($b \simeq 80\degr$) of the quasar. According to Stark et al. (\cite{sta}),
$E(B-V) \simeq 0.035$. With a galactic extinction curve this corresponds 
to an increase in galactic extinction between $\lambda = 1210\,\mathrm{\AA}$ and
$1145\,\mathrm{\AA}$ by 4\,\%, negligible compared to the other uncertainties in determining the continuum. The final adopted continuum is shown in Fig. \ref{fig1}.
The mean optical depth is $\tau_{\ion{He}{ii}} = 2.09 \pm 0.10$ at $z=2.868\pm 0.099$.
This is comparable to $\tau = 1.88$ for $z=2.82\pm 0.05$ in Q~0302-003 (Heap et al. \cite{hea}). 
In the voids (I-V) the typical optical depth is $\sim 1.6$.
Notice, however, the discussion on the possible contribution by an optical thin Lyman limit system (next section).

\subsection{Analysis of the \ion{H}{i} and \ion{He}{ii} Ly$\alpha$ forest}

An overlay of the STIS (\ion{He}{ii}) data with the Keck/HIRES data is presented
in Fig. \ref{fig2}a. The \ion{H}{i} wavelength scale has been divided by a factor of
4.00178 in order to align the \ion{H}{i} with the \ion{He}{ii} absorption. Although the
signal to noise of the STIS \ion{He}{ii} spectrum is very low ($\sim 2$), it can easily
be seen that the opacity gaps in \ion{He}{ii} (numbered I to V) correspond
to gaps in the \ion{H}{i} Ly$\alpha$ forest, so that modelling of the \ion{He}{ii}
absorption seems worthwhile. As known from high-resolution FUSE spectra
of the \ion{He}{ii} 304\,\AA\, forest in HE~2347-4342 (Kriss et al. \cite{kri},
Shull et al. \cite{shu}, Zheng et al. \cite{zhe2004b}), and in HS~1700+6416 (Reimers et al. \cite{rei2004}), the
\ion{He}{ii} forest can be reproduced roughly by scaling a model of the \ion{H}{i} Ly$\alpha$
forest using Doppler profiles with a constant column density ratio $\eta = N(\ion{He}{ii})/N(\ion{H}{i})$. The Ly$\alpha$ forest model is represented by a list
of wavelengths, broadening parameters $b$ and column densities $N_{\ion{H}{i}}$ for
all relevant Ly$\alpha$ lines. Metal lines have been identified and removed
from the \ion{H}{i} Ly$\alpha$ forest spectrum first. 
In a first approximation, we assume $\eta = 80$ which is close to the
mean value found, e.g., for HE~2347-4342 (Kriss et al. \cite{kri}) and HS~1700+6416 (Reimers et al. \cite{rei2004}), and adopt pure turbulent broadening.
Fig. \ref{fig2}b shows this synthesized spectrum in comparison with the observed
spectrum. 
However, in the voids I to V the model \ion{He}{ii} spectrum predicts much less \ion{He}{ii} opacity than
observed even for $\eta = 1000$. 
Computing a $\chi^2$, we find the probability that the models are correct to be $3\cdot 10^{-7}$ for $\eta = 80$ and $10^{-3}$ for $\eta = 1000$, respectively.
The reason for the discrepancy might be that with the
$S/N$ of $\sim 40$ of the Keck/HIRES spectrum, the low cutoff of the Ly$\alpha$ forest
is at $\log N_{\ion{H}{i}} \simeq 11.6$, while higher $S/N$ spectra allow a better placement of the continuum and show that the Ly$\alpha$ forest
extends to $\log N_{\ion{H}{i}} = 11$ (e.g. Kirkman \& Tytler \cite{kir1997}). Assuming a power law
$dn/dN = A\cdot N^{-{\beta}}$ according to Kirkman \& Tytler (\cite{kir1997}), we can predict the
number of Ly$\alpha$ lines below our cutoff. We have performed a number of numerical
experiments distributing the weak Ly$\alpha$ lines ($11.0 \leq \log N \leq 11.6$) in
different ways and assuming $\eta = 300$ to $\eta = 1000$.
It turns out, that the \ion{He}{ii} opacity in the gaps I and III-V cannot be explained in this manner, even with extreme assumptions ($\eta = 1000$). 
One possible conclusion is that the intergalactic medium in the range $2.77 \leq z \leq 2.97$ in this particular
line of sight is not yet fully reionized in \ion{He}{ii}. 
The reason might be that we are in a
strongly overdense region where \ion{He}{ii} reionisation is further delayed. This is consistent
with the observation, that Q~1157+3143 has a particularly rich metal absorption spectrum.
Ganguly et al. (\cite{gan}) found on low resolution spectra 7 \ion{C}{iv} doublets between $z = 2.68$ and
$z = 2.97$, compared to two expected and suggested superclustering at $z = 3$ as a possible reason.
We confirm this finding. In the redshift range shown in Fig. \ref{fig2}a, there
are four strong metal absorption systems associated with high column density
Ly$\alpha$ and \ion{He}{ii} absorption: $z = 2.94, 2.875, 2.84$ and $2.77$.
The 2.94 and 2.77 systems are Lyman limit systems with complex metal line absorption typically spread over a large velocity range of $350\,\mathrm{km\,s}^{-1}$
($z = 2.94$) and $250\,\mathrm{km\,s}^{-1}$ ($z = 2.77$). 
The $z = 2.875$ system has at least 3 \ion{C}{iv}
components (also \ion{Si}{iv}, \ion{Si}{iii}). The 2.84 system consists of 2 close \ion{C}{iv} pairs separated by $80\,\mathrm{km\,s}^{-1}$.
The sequence of Ly$\alpha$ + \ion{He}{ii} ``voids'' alternating with strong
\ion{H}{i} + \ion{He}{ii} + metal line absorption indeed reminds us of the filament - void
structure of the IGM. The ``void'' sizes are typically several Mpc, e.g. void II between 'filaments' $z = 2.84$ and $2.875$ has a size of $\sim 5.6\,\mathrm{Mpc}$ (comoving).  

\begin{figure}
  \centering
  \resizebox{\hsize}{!}{\includegraphics[bb=35 505 530 770,clip=]{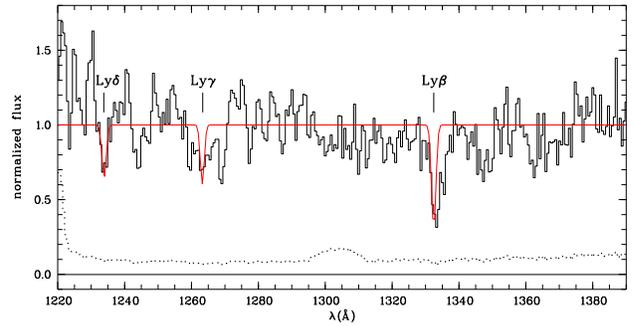}}
  \caption{Normalized STIS spectrum in comparison with Ly$\beta$ to Ly$\delta$ of a model Lyman limit system at $z = 0.299$ with $\log N(\ion{H}{i}) = 17.0$ and $b = 33\,\mathrm{km\,s}^{-1}$.
  }
  \label{fig4}
\end{figure}

We briefly discuss an alternative explanation for the observed high opacity in the \ion{He}{ii} forest which avoids the need for very high $\eta$-values and/or a contribution due to incomplete reionization as late as $z=2.8$.
Assuming that a low redshift Lyman limit system is responsible for the reduction of the flux in the opacity gaps I -- V, its optical depth must be of the order of unity at 1160\,\AA, and its redshift $z<0.33$ according to its nonvisibility above 1220\,\AA. 
As shown in Fig. \ref{fig4}, there is some observational evidence for a Lyman series at $z \approx 0.3$ visible in Ly$\beta$ (blended with IS \ion{C}{ii} 1335\,\AA), Ly$\gamma$ and Ly$\delta$. 
The strong line at 1270\,\AA\, could be \ion{C}{iii} 977\,\AA\, at $z \approx 0.3$.
Ly$\alpha$ is also visible but at such low $S/N$ that modelling is meaningless.
The calculated $z \approx 0.3$ Lyman limit system model shown in Fig. \ref{fig4} assumes $\log N(\ion{H}{i}) = 17.0$ and $b = 33\,\mathrm{km\,s}^{-1}$, where the latter is the best fit for the given column density. 
We conclude that this model is consistent with the observations.
This model shows that the $n=12$ and $n=13$ Lyman lines would cause a depression of roughly 50\,\% at 1190\,\AA\, which explains most of the "missing opacity" in the opacity gap I (Fig. \ref{fig2}a).
Some additional absorption must be due to IS \ion{Si}{ii} 1190/1193, since QSO 1157+3143 does show weak IS \ion{Ca}{ii} lines in our Keck spectrum.
The model predicts that for $\lambda< 1184\,\mathrm{\AA}$ the discrepancy between the predicted and the observed \ion{He}{ii} spectrum can be explained by the postulated $z \approx 0.3$ LLS.
We have also studied a possible contribution by other metal line systems and found no significant contribution among the known 21 metal line systems.
Even in case of HS~1700+6416 with its extremely rich metal line spectrum (at least 7 Lyman limit systems, Reimers et al. \cite{rei1992}) only 13\,\% of the \ion{He}{ii} forest features are affected (Reimers et al. \cite{rei2004}).

We also notice that around $z = 2.88$ (region A in Fig. \ref{fig2}b) there is strong \ion{He}{ii} absorption
without corresponding \ion{H}{i} absorption, similar to the not yet reionized patches seen in HE~2347-4342
(Reimers et al. 1997).
Compared to other lines of sight. in particular that of HE~2347-4342, the \ion{He}{ii} reionization is less complete in the voids (I-V).
Numerical simulations (e.g. Fardal et al. \cite{far}) show that indeed the intensity fluctuations in the \ion{He}{ii} continuum are much larger than in the \ion{H}{i} continuum.
The reason is that the \ion{He}{ii} attenuation sphere is smaller and the QSOs with hard (flat) spectra are rarer than the steep-spectrum QSOs (cf. Shull et al. \cite{shu}).
QSO~1157+3143 itself is probably the nearest quasar with a steep ($\alpha = 2$), soft spectrum which may explain the nonvisibility of a proximity effect.

\section{Discussion}

In this paper we have analysed STIS (\ion{He}{ii}) and Keck/HIRES data in
the redshift range $z = 2.75$ to $2.97$ toward QSO 1157+3143. From previous
\ion{He}{ii} observations, in particular of HE~2347-4342 (Reimers et al. \cite{rei1997},
Smette et al. \cite{sme}, Kriss et al. \cite{kri}) and Q~0302-003 (Heap et al. \cite{hea}), we know
that in this redshift range we observe the transition from 'black'
absorption troughs ($z > 3$) through patchy \ion{He}{ii} absorption (alternating
black troughs and optically thin \ion{He}{ii} absorption, $2.8 \leq z \leq 3$)
to a more homogeneously ionized \ion{He}{ii} Ly$\alpha$ forest which is formed in the reionized phase.
This has been interpreted as the final breakthrough and rapid completion of
\ion{He}{ii} reionisation, where the deep absorption troughs seen in HE~2347-4342
between $z = 2.83$ and $2.89$ are not yet reionized patches.
The \ion{He}{ii} data of QSO 1157+3143 confirm this picture in that such troughs
(remnant \ion{He}{ii} regions between \ion{He}{iii} regions) are also seen at $z = 2.88$
and $z = 2.94$.
In addition, attempts to model the \ion{He}{ii} absorption in the ``voids'' I to V by means of the
Ly$\alpha$ forest with column densities as low as $\log N = 11.0$, have been largely unsuccessful.
We suspect, that \ion{He}{ii} reionisation is not complete even in ``voids'' due to large overdensities
of the IGM just in front of QSO 1157+3143.
The three QSOs with \ion{He}{ii} spectra covering the redshift range $z = 2.75$
to $3.0$ map the phase where the remaining \ion{He}{ii} blobs between the expanding
\ion{He}{iii} regions finally become optically thin.
In Table \ref{tab2} we compare the three lines of sight.

\begin{table}
  \caption[]{
  }
  \label{tab2}
  \begin{center}
  \begin{tabular}{lcc}
    \hline
    \noalign{\smallskip}
     & Lowest $z$  with               & Highest $z$ with \\
     & $\tau(\ion{He}{ii}) >> 1$ with & remnant \\
     & no corresponding \ion{H}{i}    & QSO flux \\
    \noalign{\smallskip}
    \hline
    \noalign{\smallskip}
    HE~2347-4342  & 2.83 & 2.87 \\
    Q~0302-003    & 2.88 & \ \ 2.90 $^{\mathrm{a}}$\\
    QSO~1157+3143 & 2.88 & 2.92 \\ 
    \noalign{\smallskip}
    \hline
  \end{tabular}
  \end{center}
\begin{list}{}{}
  \item[$^{\mathrm{a}}$] Except the $z = 3.05$ low opacity feature which is due to the transverse proximity effect by a QSO close to the line of sight (Jakobsen et al. \cite{jak2003}).\\
\end{list}
\end{table}

The results shown in Table \ref{tab2} indeed give a consistent picture of the final
phase of \ion{He}{ii} reionisation. The transition from optically thick \ion{He}{ii}
absorption remaining from the not yet reionized phase ($z \geq 3.5$)
to the phase where reionisation is complete occurs between $z = 2.9$ and $z = 2.8$. The relatively precise measurement of the completion of an important
cosmic epoch has to be reproduced by future theoretical models of \ion{He}{ii}
reionisation. However, we also note that cosmic variance cannot be neglected.

A further new result seen so far only in QSO 1157+3143 is the clear
indication of a filament/void structure in the intergalactic 
\ion{He}{ii} 304\,\AA\, opacity. That this can be seen in spite of the poor
quality of the \ion{He}{ii} data is due to the fact that the line of sight of
QSO 1157+3143 is particularly rich in strong metal line systems in the
redshift range $2.7 \leq z \leq 2.97$.
Between these high density structures also seen in the Ly$\alpha$ forest,
there are a number of clear voids (structures I - V in Fig. \ref{fig2}a) in the
Ly$\alpha$ forest, i.e. the ``density contrast'' in the large scale
structure is particulary strong just in front of this QSO.
Without the $z \approx 0.3$ LLS absorption, the filament/void structure would have been more pronounced in the \ion{He}{ii} forest.

\section{Summary}

\begin{figure*}
  \centering
  \resizebox{\hsize}{!}{\includegraphics[bb=45 265 525 755,clip=]{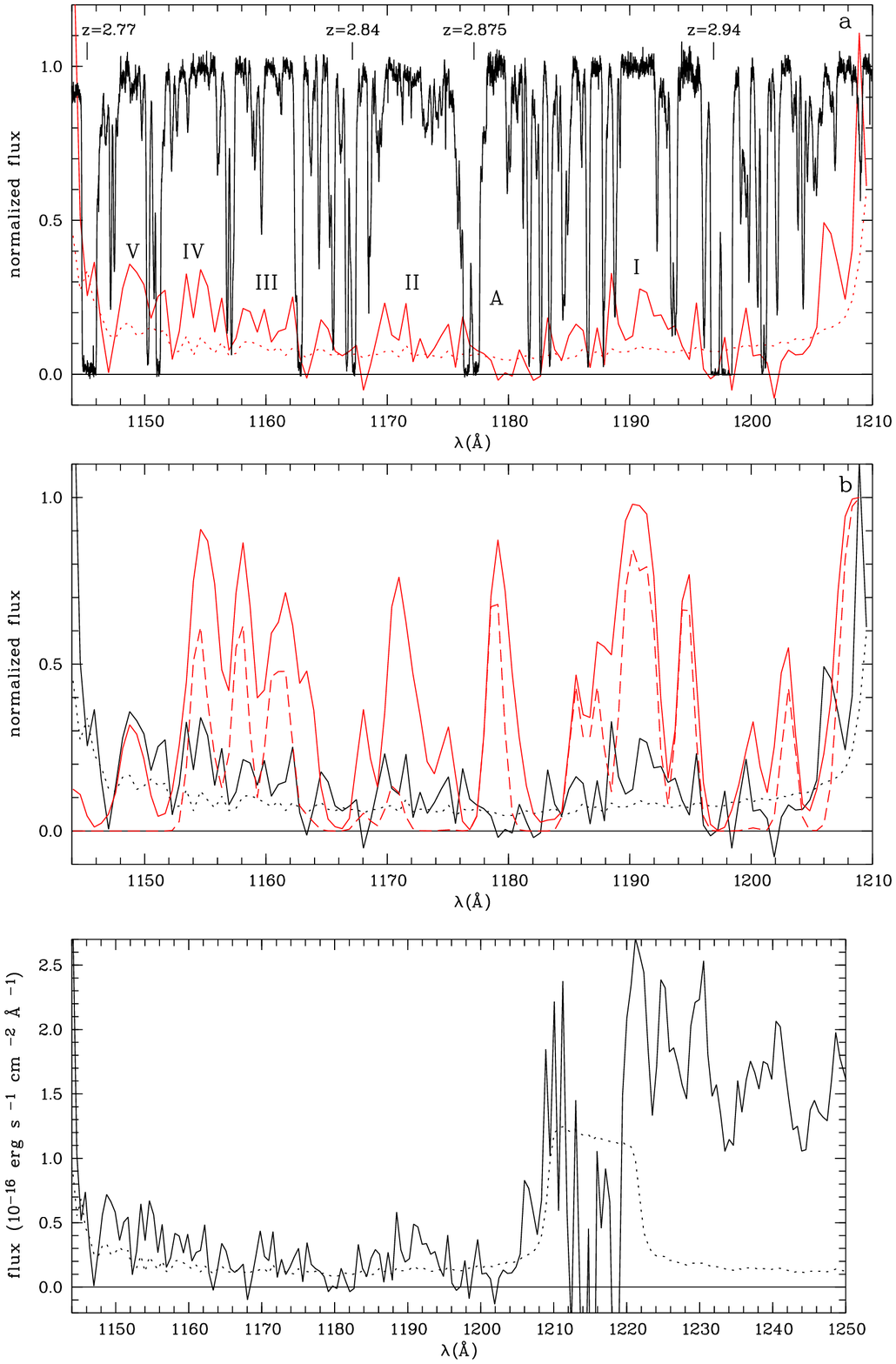}}
  \caption{
    {\bf a} (upper). Overlay of the Ly$\alpha$ absorption of \ion{He}{ii} (from HST/STIS, bottom) and of \ion{H}{i} (Keck/HIRES, top). Wavelengths for the \ion{H}{i} data are divided by 4.00178 to align them with the \ion{He}{ii} data. The strong Ly$\alpha$ lines belonging to strong metal absorption systems (cf. text) are marked.
    {\bf b} (lower). Overlay of the \ion{He}{ii} Ly$\alpha$ absorption (thick) and of the \ion{He}{ii} model absorption (thin: $\eta = 80$, dashed: $\eta = 1000$). The mean uncertainty per pixel is 0.03 in case of $\eta = 80$ and 0.005 for $\eta = 1000$, respectively. The opacity gaps (voids) I to V and the lowest $z$ black trough (A) are marked. The dotted line represents the error of the STIS data.
  }
  \label{fig2}
\end{figure*}

This paper contributes with QSO 1157+3143 a new line of sight for studying
intergalactic \ion{He}{ii} absorption.
We have obtained low resolution HST/STIS spectra of the \ion{He}{ii} 304\,\AA\,
forest and high-resolution Keck/HIRES spectra of the corresponding \ion{H}{i}
Ly$\alpha$ forest.
Both in \ion{He}{ii}, \ion{H}{i} and in strong metal line systems the line of sight
towards QSO 1157+3143 shows for $2.77 < z < 2.97$ an alternating
sequence of voids (\ion{He}{ii} and \ion{H}{i} absorption weak) and filaments (\ion{He}{ii}, \ion{H}{i} strong, complex \ion{C}{iv} etc. lines).
Due to the rich, high-contrast large scale structure in front of 
QSO 1157+3143, this phenomenon has been seen for the first time in \ion{He}{ii} in
this line of sight. We also observe the patchiness of the \ion{He}{ii} opacity seen first in
HE~2347-4342 (Reimers et al. \cite{rei1997}), i.e. remnant strong \ion{He}{ii} opacity spots without \ion{H}{i}
counterparts like region A (Fig. \ref{fig2}a) at $z = 2.88$, further evidence for not yet completed
\ion{He}{ii} reionisation. Even in the ``voids'' (regions I-V in Fig. \ref{fig2}a), reionisation of \ion{He}{ii}
is probably not complete, since the \ion{He}{ii} opacity cannot be modelled by means of the
Ly$\alpha$ forest alone. Together with HE~2347-4342 and Q~0302-003, the line of sight
towards QSO 1157+3143 provides further evidence that the transition between optically thick \ion{He}{ii}
absorption left over from the not reionized phase to reionisation being completed occurs between
$z = 3$ and $z = 2.7$.

\begin{acknowledgements}
We thank the referee for valuable comments which helped to improve the paper.
This work has been supported by the Verbundforschung of the BMBF/DLR under Grant No. 50 OR 9911 1.
DT and DK were supported by STScI grant GO-9350, and in part by NSF grant AST-0098731 and NASA grant NAG5-13113.
\end{acknowledgements}

{}

\end{document}